# Facile and time-resolved chemical growth of nanoporous Ca$_x$CoO$_2$ thin films for flexible and thermoelectric applications

Tridib Kumar Sinha[1,2], Jinho Lee[3], Jin Kuk Kim[2], Samit K. Ray[1], Biplab Paul[4*]

[1]*Department of Physics, Indian Institute of Technology, Kharagpur 721302, India*
[2]*Department of Materials Engineering and Convergence Technology & [3]The Research Institute of Natural Science and Department of Physics Education, Gyeongsang National University, Jinju 52828, South Korea*
[4]*Thin Film Physics Division, Department of Physics, Chemistry, and Biology (IFM),* Linköping *University, SE-581 83* Linköping, *Sweden*

*\*Corresponding Author:* biplab.paul@liu.se

**Abstract**

Ca$_x$CoO$_2$ thin films can be promising for widespread flexible thermoelectric applications in a wide temperature range from room-temperature self-powered wearable applications (by harvesting power from body heat) to energy harvesting from hot surfaces (e.g., hot pipes) if a cost-effective and facile growth technique is developed. Here, we demonstrate a time resolved, facile and ligand-free soft chemical method for the growth of nanoporous Ca$_{0.35}$CoO$_2$ thin films on sapphire and mica substrates from a water-based precursor ink, composed of in-situ prepared Ca$^{2+}$-DMF and Co$^{2+}$-DMF complexes. Mica serves as flexible substrate as well as sacrificial layer for film transfer. The grown films are oriented and can sustain bending stress until a bending radius of 15 mm. Despite the presence of nanopores, the power factor of Ca$_{0.35}$CoO$_2$ film is found to be as high as $0.50 \times 10^{-4}$ Wm$^{-1}$K$^{-2}$ near room temperature. The present technique, being simple and fast to be potentially suitable for cost-effective industrial upscaling.

## 1. Introduction

Since the discovery of single crystalline Na$_{0.50}$CoO$_2$ exhibiting a large Seebeck coefficient and low electrical resistivity in 1997 [1] extensive research interests have been paid to develop new layered cobaltates (alias cobalt oxides) of significant thermoelectric (TE) performance. In this regard, Na$_{0.7}$CoO$_2$ is reported to exhibit a high power factor (PF=$S^2/\rho$, where $S$ is the Seebeck coefficient, $\rho$ is the electrical resistivity) similar to the standard Bi$_2$Te$_3$ [1], however with one drawback that Na$_{0.7}$CoO$_2$ due to its poor chemical stability cannot offer sustainable performance, particularly, at elevated temperatures. Many sustainable layered cobaltates, such as $Ca_xCoO_2$ [2,3], $Sr_xCoO_2$ [4,5] and $Ln_{0.30}CoO_2$ (Ln = La, Pr and Nd) [6,7] have been developed simply by exchanging the $Na^+$ ions with other divalent or trivalent cations from $Na_xCoO_2$ precursor. Among all, calcium cobaltates (e.g. $Ca_xCoO_2$ and $Ca_3Co_4O_9$) has attracted significant interest because of their high TE performance, remarkable thermal and chemical stability and nontoxicity. In particular, thin films of these calcium cobaltates can be promising for flexible applications if the film deposition be possible on flexible substrates or otherwise transferable onto flexible platforms.

Since the thermal conductivity of $Ca_xCoO_2$ is higher than $Ca_3Co_4O_9$, the latter has been more investigated for high temperature thermoelectric applications. For high output from a thermoelectric converter, its constituent leg materials need to have high power factor [8]. In particular, achieving high power factor is more important than efficiency for low power applications (such as wearable applications). The $Ca_{0.35}CoO_2$ is reported to exhibit a high power factor of 0.9 $mWm^{-1}K^{-2}$ at 300 K [9], which is several times higher than the values reported for misfit layered cobaltate $Ca_3Co_4O_9$ [10-14], and remarkably higher than other layered cobaltates $A_xCoO_2$ [A = Sr, La, Pr, Nd] [4-7].

Compared to the bulk materials, thin films of these layered cobaltates are advantageous because of their low-dimension, cost-effectiveness and easy integration in different platforms for advanced applications. Although physical deposition techniques have been investigated for the growth of $Ca_xCoO_2$ thin films [15-17] the investigation on chemical solution deposition (CSD) method is still elusive. On the other hand, to address the scalability, chemical growth methods are often preferred over physical methods. Although some investigations have already been done to grow the misfit layered cobaltate thin films by chemical methods [18-20], however, with no report on $Ca_xCoO_2$. Furthermore, the challenges associated with the chemical methods are from the nonuniformity of the films [18] due to the undesired cross-reactions or non-homogenous precipitation. Panchakarla et al. reported the growth of misfit layered $Ca_3Co_4O_9$ nanotubes by base-treatment of pre-formulated calcium cobalt oxide [21] but the process seems to be complicated to obtain thin films. The growth of high quality thin film of misfit layered cobaltate has been accomplished using a water-based solutions containing the co-ordinate complex of metal ions with EDTA (ethylenediaminetetraacetic acid) and PEI (polyethileneimine) [20]. However, the processing of the complex-solution remains tedious and time consuming.

Therefore, development of alternative strategies is imperative to produce good quality $Ca_xCoO_2$ thin films in a facile and time-resolved process that also allow their easy integration in multiple modules. Here, we demonstrate a ligand-free water-based CSD method for the growth of $Ca_{0.35}CoO_2$ thin films on sapphire and flexible mica substrate. Muscovite mica, because of its inherently layered structure, where aluminosilicate layers are loosely bound by the boundary layer of potassium (K+) ions, can serve as the flexible substrate [22]. Furthermore, it can sustain high processing temperature of 700 ºC. The film growth technique being simple and low cost is potentially suitable for industrial upscaling. TE performance of the films is characterized in terms of their power factor. The nanoporous structure of the film is found not to affect its power factor, however, the presence of pores of nanometer length-scale in the film can be expected to reduce thermal conductivity of the film as compared to the pristine bulk value [23,24].

## 2.  Experimental section

### General consideration

Metal ions such as $Ca^{2+}$, $Co^{2+}$, etc. are very much susceptible for complexation with different polar solvents such as DMF [25-34]. The DMF is cheaper and easily available as compared to other ligands, e.g., EDTA, PEI, used for the growth of cobaltate thin films [20]. It is also non-hazardous [35] and recyclable.

### Formulation of DMF-complexes with $Ca^{2+}$ and $Co^{2+}$

In this work, hydrated acetate salts of $Ca^{2+}$ $(Ca(CH_3COO)_2.H_2O)$ and $Co^{2+}$ $(Co(CH_3COO)_2.H_2O)$ were taken as the precursor materials which were easily soluble in DMF. The DMF complexes of the metal ions namely, $Ca^{2+}$-DMF and $Co^{2+}$-DMF were prepared simply by adding the salts separately in a molar ratio 3:4 into the excess DMF, followed by stirring for 30 min at 60°C. After cooling at room temperature white precipitate of $Ca^{2+}$-DMF and violet precipitate of $Co^{2+}$-DMF appeared (as shown in Figure 1 (a)), which were re-precipitated (for 4-times) from hot DMF. Here, the precursor salts and DMF were procured from Sigma Aldrich.

### Formulation of precursor thermoelectric ink

The precipitates were mixed together by dissolving in excess of DMF followed by stirring for 1 hour at 90°C. After cooling at room temperature, a purple color precipitate was obtained (as shown in Figure 1) which was re-precipitated from hot DMF (for 4-times) to remove the unwanted impurities. After washing with ether, the precipitate was dried at 60 °C to obtain a pink brown color precursor solid (as shown in extreme left of Figure 1(b)), which was easily soluble in de-ionized (DI) water to obtain a stable homogenous solution of bluish precursor ink having solid content of 10mg/mL.

**Fabrication of thermoelectric samples**

For the investigation thin films of $Ca_{0.35}CoO_2$ were grown by CSD method on sapphire and mica substrates. The ink was simply drop-casted onto the top-surfaces of cleaned substrates and heated to 700°C for 10 min in ambient condition. Subsequently, black-colored thin film of $Ca_{0.35}CoO_2$ appeared on the substrates.

**Characterization**

The crystal structure and morphology of the films were characterized by $\theta$ –$2\theta$ XRD analyses using monochromatic Cu K$\alpha$ radiation ($\lambda = 1.5406$ Å) and scanning electron microscopy (SEM, LEO 1550 Gemini). The $\theta$–$2\theta$ XRD scans were performed with a Philips PW 1820 diffractometer. Compositional analyses of the films were performed by energy dispersive X-ray spectroscopy (EDS) with an accuracy of ±5%. In-plane electrical resistivity and Seebeck coefficient were simultaneously measured as a function of temperature using an ULVAC-RIKO ZEM3 system.

## 3.   Results and Discussion

Figure 1 shows the different steps for the growth of $Ca_{0.35}CoO_2$ thin-films on mica and sapphire substrates, which is partly discussed before in the experimental section. Figure 1(a) and (b) show the step-by-step formulation of the water-based precursor thermoelectric ink from the corresponding solid mixture of $Ca^{2+}$- and $Co^{2+}$-DMF complexes. The DMF-complexes of the metal ions have been formulated by adding their acetate salts (i.e., Ca-acetate and Co-acetate) in hot (60 ºC) DMF followed by cooling at room temperature. The DMF is a highly susceptible material to form complexes with the metal ions through its amide functionality (as schematically shown in Figure 1 (a)). Although, the precursor salts are soluble in hot DMF, the precipitation was observed during cooling at room temperature, which is indicative to the formation of DMF-complexes of $Ca^{2+}$ and $Co^{2+}$.

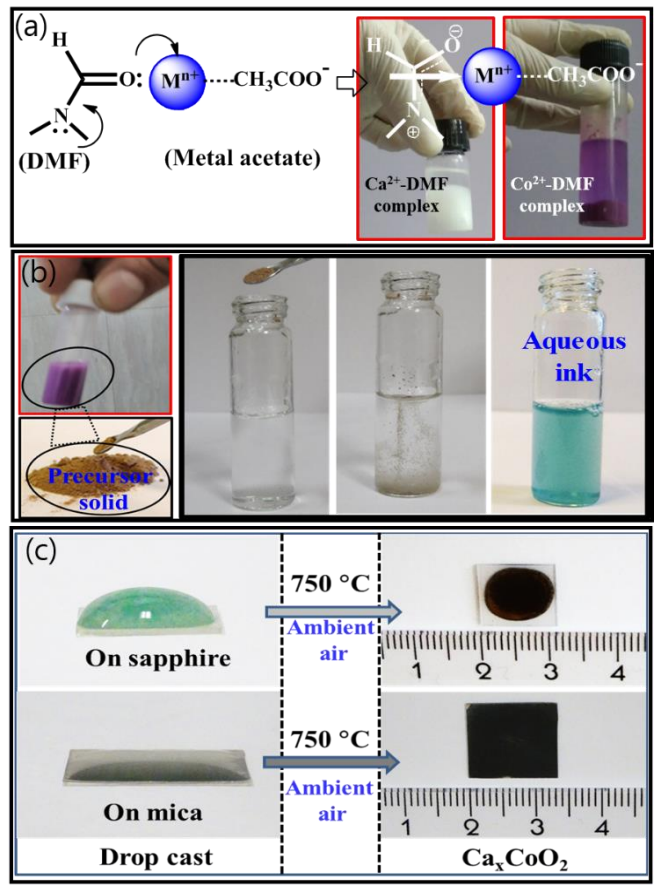

**Figure 1**. (a) Formation of $Ca^{2+}$-DMF and $Co^{2+}$-DMF complexes as white and violet precipitates respectively, (b) formation of pink brown precursor solid and the preparation of aqueous thermoelectric ink, (c) chemically grown $Ca_{0.35}CoO_2$ thin film on sapphire and mica substrates by simple drop casting of the precursor ink followed by annealing at 700 ºC in air.

During formation of the DMF-complexes, there is a possibility of partial substitution of the acetate ions (-$COOCH_3^-$) by the DMF. Consequently, the DMF-complexes are thought to behave like zwitterions. During mixing of the DMF-complexes in excess DMF with heating and vigorous stirring, the zwitterions electrostatically interacts each other, resulting in formation of electrostatically balanced precipitate of the complex-conjugates, which was dried to obtain the pink brown precursor solid (as shown in Figure 1 (b)). Because of abundant ionic characteristics of the solid, it likely to be soluble in polar solvent. Water being the most environment friendly polar solvent, here, we have demonstrated the formulation of highly stable (stability was noticed for 60 days) aqueous thermoelectric ink.

The aqueous solution was drop-casted on two different substrates, sapphire and mica

(Figure 1(b)). After heat treatment at 700 °C in ambient atmosphere, dark films were obtained. The $Ca_{0.35}CoO_2$ film on mica substrate was found to be more uniform as compared that on sapphire substrate. This is because, angle of contact between solution and hydrophilic mica [36,37] is much smaller than between solution and sapphire, which results in uniform spreading of the solution over mica substrate. The formation of $Ca_{0.35}CoO_2$ thin films was repeatedly observed. During mixing of the DMF-complexes under vigorous stirring, the $Co^{2+}$-DMF complex (because of its smaller size) will occupy within the interstitial spaces of $Ca^{2+}$-DMF through the favorable electrostatic interaction among the active groups in the periphery of metal ions, resulting in-situ electrically and stoichiometrically balanced thermoelectric precursor solid (i.e., $Ca^{2+}/Co^{2+}$-DMF mixed complex). During heat treatment, the coordinated DMF reduces the $Co^{2+}$ to $Co^0$ nanoparticles, which in presence of abundant oxygen transforms to $CoO_2$. Plausible reaction mechanism is shown in Scheme 1. On application of elevated temperature in air, the DMF degrades to produce $H_2O$, $CO_2$, $N_2$, $NO_x$, etc., resulting porosity in the film [38,39]. $Ca^{2+}$-DMF is ionized at high temperature and forms $Ca_{0.35}CoO_2$.

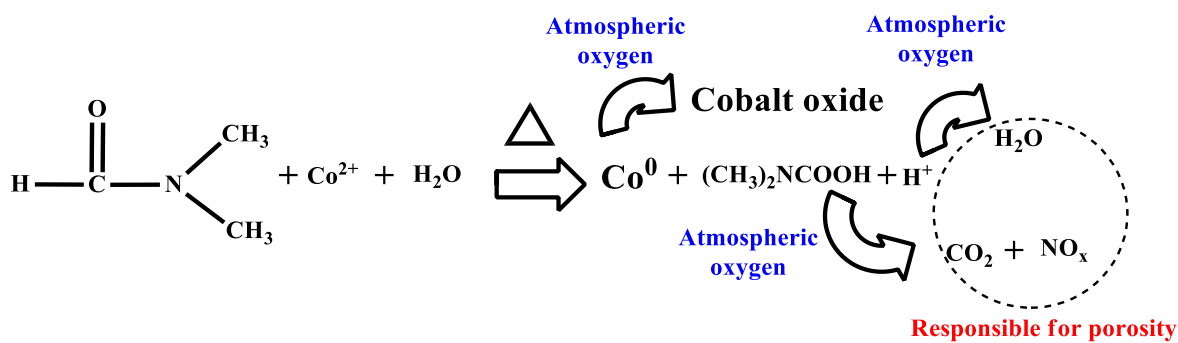

**Scheme 1**. Reduction of $Co^{2+}$ by DMF and formation of nanoporous $Ca_{0.35}CoO_2$ thin films.

Figure 2 (a) and (b) are the XRD scan of the films $Ca_{0.35}CoO_2$-on-mica and $Ca_{0.35}CoO_2$-on-sapphire, respectively. Both the figures show $Ca_{0.35}CoO_2$ 001 and 002 peaks. The peaks from no other planes, except (00$l$) planes of $Ca_{0.35}CoO_2$ indicate the $c$-axis orientation of both the films. Apart from $Ca_{0.35}CoO_2$, the peaks from mica and sapphire substrates are also visible in Figure 2 (a) and (b) respectively. The d-spacing for both the films is calculated to be 5.437 Å, which nearly matches with the value reported elsewhere [40].

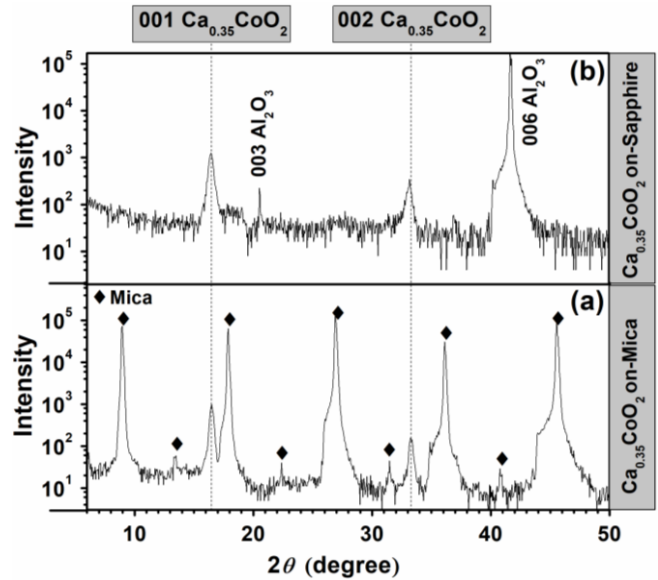

**Figure 2.** XRD pattern of (a) $Ca_{0.35}CoO_2$ film on mica substrate, (b) $Ca_{0.35}CoO_2$ film on sapphire substrate.

Figure 3 (a) shows a typical SEM image of $Ca_{0.35}CoO_2$ film deposited on sapphire substrate. Random distribution of triangular grains of $Ca_{0.35}CoO_2$ is visible in Figure 3 (a). The presence of nanopores in the film is indicated by the white open circles in Figure 3 (a). Figure 3 (b) shows the magnified SEM image of the film. Inset of Figure 3 (b) shows the polygonal shape of some fine grains of $Ca_{0.35}CoO_2$, typically of dimension less than 200 nm. Figure 3 (c) shows a typical SEM image of $Ca_{0.35}CoO_2$ film deposited on mica substrate. Random distribution of nanopores of irregular shapes is clearly visible in Figure 3 (c). Figure 3 (d) shows the magnified SEM image of the same film. Relatively smooth surface of the film and the absence of any larger grains above 200 nm is evident from the figure. Inset of Figure 3 (d) shows the polygonal shape of the grains of $Ca_{0.35}CoO_2$.

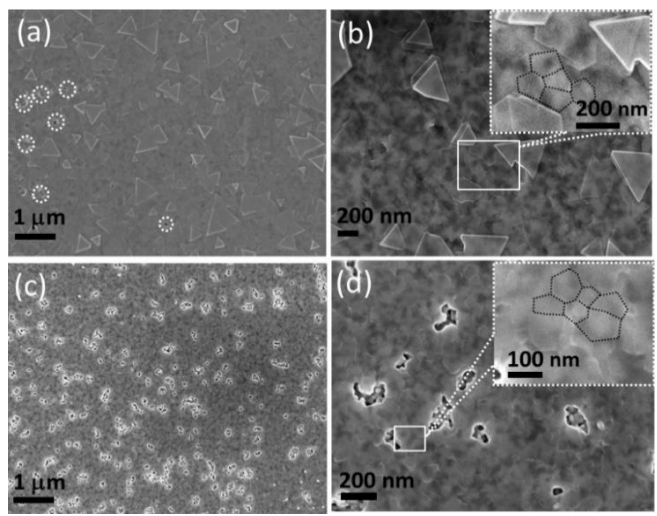

**Figure 3.** (a) SEM micrograph of $Ca_{0.35}CoO_2$ film on sapphire substrate. (b) Magnified image of a small portion of the film $Ca_{0.35}CoO_2$-on-sapphire; inset shows the polygonal shape of the grains in the film. mica substrate. (c) SEM micrograph of $Ca_{0.35}CoO_2$ film on mica substrate. (d) Magnified image of a small portion of the film $Ca_{0.35}CoO_2$-on-mica substrate; inset shows the polygonal shape of the fine grains in the film.

Seemingly flat surface of both the films is attributed to the out-of-plane orientation of the film, which is in consistent with the observation by XRD. The growth of oriented films on both the substrates is intrinsically attributed to the effect of self-assembly during annealing, where the driving force is the external stress due to solvent evaporation [41]. The higher porosity of $Ca_{0.35}CoO_2$ film on mica substrate as compared to that on sapphire substrate might be due to the slow evaporation of the surface absorbed water in the more hygroscopic mica during the formation of the film from the precursor ink [42].

From the above study it is conjectured that the present CSD method can be applicable to grow $Ca_{0.35}CoO_2$ thin films on arbitrary substrates, who sustain high annealing temperature 700 ºC. The porosity of the films can also be controlled by preferential selection of the substrate with desired hygroscopic properties.

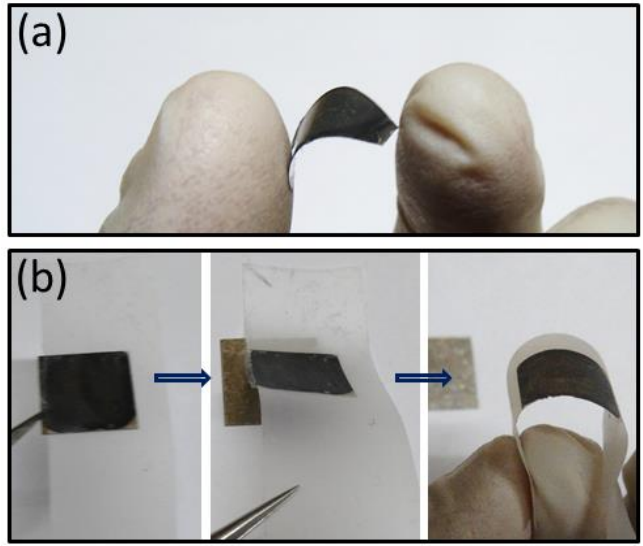

**Figure 4.** (a) Image of thin flexible film $Ca_{0.35}CoO_2$-on-mica. (b) Demonstration of the film transfer onto a sticky tape.

Figure 4 (a) shows a typical optical image of bended $Ca_{0.35}CoO_2$ film of thickness ~300 nm on mica substrate (25 µm thickness). The film is bendable to a bending radius of 15 mm without developing any crack in the film, as confirmed by optical microscopy. In Figure 4 (b) the transferability of the film is demonstrated by transferring the film from the parent mica substrate onto a sticky tape. The top surface of the film is first adhered to a sticky tape and then the film is transferred by stripping, i.e. the mica substrate can act as sacrificial layer.

Figure 5 (a) shows temperature dependent electrical resistivity of $Ca_{0.35}CoO_2$ films on sapphire (red circles) and mica (black open circles) substrates. The room temperature resistivity values are 15.6, and 14.6 m$\Omega$cm for the films $Ca_{0.35}CoO_2$-on-sapphire, and $Ca_{0.35}CoO_2$-on-mica, respectively, which higher than the values for $Ca_{0.35}CoO_2$ reported elsewhere [16,43], however, comparable to the reported values for undoped $Ca_3Co_4O_9$ thin films [44,45]. Even with nanoporous structure the electrical resistivity of $Ca_{0.35}CoO_2$-on-mica is slightly lower than the film $Ca_{0.35}CoO_2$-on-sapphire, which is attributed to the better crystalline quality of former than later. This is apparent that the nanopores in the film $Ca_{0.35}CoO_2$-on-mica do not act as scattering centre for the charge carriers, which might be due to the higher interpore separation than the electronic mean free path. However, thermal conductivity of the film is anticipated to be reduced, due to the enhanced scattering of phonons by nanopores. This anticipation is since the phononic mean free path is one order of magnitude higher than the electronic mean free path, and by controlling the characteristic length scale of the nanoporous structure within the range of electronic and phononic mean free path the phonons can be selectively scattered, but without hampering the electronic transport [23,46]. Figure 4 (b) shows the temperature dependent Seebeck coefficient of the films. The room temperature Seebeck coefficient values are 78 and 88 $\mu$V/K for the films $Ca_{0.35}CoO_2$-on-sapphire, and $Ca_{0.35}CoO_2$-on-mica, respectively, which are comparable to the value reported for single crystalline $Ca_{0.33}CoO_2$ [9], however, sufficiently higher than sputter deposited $Ca_xCoO_2$ thin films reported elsewhere [16]. The Seebeck coefficient varies with temperature in a similar manner as electrical resistivity. Figure 4c shows the temperature dependent power factor of the grown films. The room temperature power factor values are 0.39 and $0.50 \times 10^{-4}$ $Wm^{-1}K^{-2}$ for the film $Ca_{0.35}CoO_2$-on-sapphire, and $Ca_{0.35}CoO_2$-on-mica, respectively, which are interestingly comparable to the reported values for undoped $Ca_3Co_4O_9$ thin films [44,45].

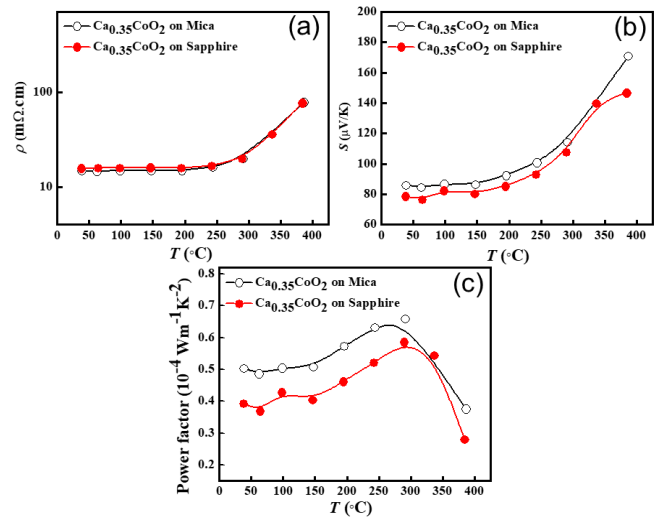

**Figure 5.** Temperature dependent (a) electrical resistivity, (b) Seebeck coefficient, and (c) power factor of $Ca_{0.35}CoO_2$ films grown on sapphire and mica substrates.

The slightly higher power factor of the film $Ca_{0.35}CoO_2$-on-mica as compared to $Ca_{0.35}CoO_2$-on-sapphire, is attributed to the reduced electrical resistivity of the former than later. To investigate the effect of bending stress on thermoelectric performance the film $Ca_{0.35}CoO_2$-on-mica was subjected to repeated bending, 100 times, and then Seebeck measurements were performed. No notable changes in the values of Seebeck coefficient and electrical resistivity were found; whatever little variation in the values are within the error limit specified for the measuring instrument. It is noteworthy that the present films are undoped and further improvement of power factor is still possible upon doping.

## 4. Conclusions

We have demonstrated a facile and eco-friendly chemical route to prepare the aqueous thermoelectric precursor ink for the growth of $Co_{0.35}CoO_2$ thin films on sapphire and mica substrates. The growth of nanoporous $Co_{0.35}CoO_2$ thin film has been realized, following the steps of drop casting of precursor ink and then annealing at 700 ºC for only 10 min in air. The film deposition is expected to be possible on arbitrary substrates, who sustain high annealing temperature 700 ºC. The formation of nanopores in the film is attributed to the degradation of DMF, producing $H_2O$, $CO_2$, $N_2$, $NO_x$, etc., during the heat treatment. The higher porosity of the film $Co_{0.35}CoO_2$-on-mica as compared to the film $Co_{0.35}CoO_2$-on-sapphire is attributed to the slow evaporation of the surface absorbed water in the more hygroscopic mica during the growth of the film. It is anticipated that the control over the porosity of the films is possible by preferential selection of substrates with desired hygroscopic properties. Despite the high porosity the power factor of the film $Co_{0.35}CoO_2$-on-mica is found to be higher than the film $Co_{0.35}CoO_2$-on-sapphire, which is attributed to the better crystalline quality of the former than later. Thermoelectric properties of the film $Co_{0.35}CoO_2$-on-mica are found to be unaltered even after the repeated bending, and hence they can be useful for flexible thermoelectric applications. The room temperature value of power factor of the flexible film is obtained as $0.50 \times 10^{-4}$ $Wm^{-1}K^{-2}$, which is comparable to the reported values for undoped $Ca_3Co_4O_9$ thin films. Further improvement in power factor is still possible by optimal doping. The $Co_{0.35}CoO_2$ film is also transferable from primary mica substrate to arbitrary polymer platform by simple dry transfer. The present thin film growth technique from thermoelectric precursor ink, being simple and less time consuming, is potentially suitable for industrial upscaling.

## Acknowledgements

This research was supported by Science and Engineering Research Board (SERB/1759/2014-15), DST "GPU", Government of India, Basic Science Research Program through the National Research Foundation of Korea (NRF) funded by the Ministry of Education (2016R1D1A1B03931391), and funding from the Åforsk foundation. P. Eklund (Linköping University) is acknowledged for his critical reading of the manuscript and additional funding through the Swedish Government Strategic Research Area in Materials Science on Functional Materials at Linköping University (Faculty Grant SFO-Mat-LiU No. 2009 00971).